**Title:**

# Liquid-liquid transition in water from first principles

**Authors:**

Thomas E. Gartner III[1,†], Pablo M. Piaggi[1], Roberto Car[1,2,3,4], Athanassios Z. Panagiotopoulos[5,*], and Pablo G. Debenedetti[5,*]

[1]Department of Chemistry, Princeton University, Princeton, NJ 08544;

[2]Department of Physics, Princeton University, Princeton, NJ 08544;

[3]Program in Applied and Computational Mathematics, Princeton University, Princeton, NJ 08544;

[4]Princeton Institute for the Science and Technology of Materials, Princeton University, Princeton, NJ 08544;

[5]Department of Chemical and Biological Engineering, Princeton University, Princeton, NJ 08544

[†]Current address: School of Chemical and Biomolecular Engineering, Georgia Institute of Technology, Atlanta, GA 30332

[*]Corresponding authors: azp@princeton.edu, pdebene@princeton.edu

**Abstract:**

A longstanding question in water research is the possibility that supercooled liquid water can undergo a liquid-liquid phase transition (LLT) into high- and low-density liquids. We used several complementary molecular simulation techniques to evaluate the possibility of an LLT in an *ab initio* neural network model of water trained on density functional theory calculations with the SCAN exchange correlation functional. We conclusively show the existence of a first-order LLT and an associated critical point in the SCAN description of water, representing the first definitive computational evidence for an LLT in water from first principles.

The idea that water may undergo a liquid-liquid phase transition (LLT) stems from a seminal work by Poole et al. [1], who, inspired by experimental observations by Mishima and coworkers that water's amorphous solid state exhibits distinct high- and low-density forms [2,3], used simulations of the empirical ST2 water model to study supercooled liquid water and proposed the existence of an LLT as a means of rationalizing their computational results. In the LLT viewpoint, if maintained in the supercooled liquid state at low temperatures and moderate positive pressures, water (which exists as a single liquid phase at ambient conditions) undergoes a phase transition into high-density liquid (HDL) and low-density liquid (LDL) phases [4,5]. This line of phase coexistence terminates in water's liquid-liquid critical point (LLCP), and critical fluctuations emanating from the LLCP along the Widom line are responsible for several of liquid water's anomalous physical properties [5,6], such as sharp increases [7,8] and eventual maxima [9,10] in isothermal compressibility and heat capacity upon cooling at ambient pressure.

Apart from providing a thermodynamic explanation for water's anomalies, LLTs are of significant scientific and engineering interest. For example, LLTs in mixtures are widely used in industrial separations processes [11], and LLTs play an increasingly scrutinized role in cellular function [12]. However, relatively few pure substances undergo an LLT, a phenomenon that largely occurs in liquids with strongly tetrahedral or network-forming character, such as phosphorous [13], sulphur [14], silicon [15], triphenyl phosphite [16], and (potentially) water [17]. In an important step toward understanding the microscopic basis for the presence/absence of an LLT in pure fluids, computational work has suggested that the stability of an LLT in tetrahedral liquids can be tuned via the softness of the interparticle interactions [18] or the angular flexibility of directional attractive interactions [18,19]. In this context, definitively categorizing the substances that exhibit an LLT and further illuminating the physical driving forces at play is an effort of both practical and fundamental importance.

Recent experiments on supercooled water have pushed closer to directly probing the LLT, preventing crystallization in the deeply supercooled liquid via rapid cooling of small water droplets [9,20] or rapid heating of water's amorphous solid phases [17,21]. While such efforts are providing ever stronger evidence consistent with the existence of an LLT, precisely locating such a transition in the temperature-pressure plane is challenging, due to the short time scales of the experiments. On the computational side, classical molecular models such as ST2, TIP4P/2005, and TIP4P/Ice have been rigorously shown to exhibit an LLT [22,23]. However, these models use simple empirical expressions to model water's intermolecular interactions, parameterized to match experimental thermophysical data, and thus cannot in principle provide definitive evidence that water itself possesses an LLT. Several other more complex models, some of which include additional levels of chemical realism such as many-body and/or polarizability effects, have also demonstrated evidence consistent with an LLCP [24-27]. Recently, the WAIL model, which captures bond flexibility and polarizability effects via fits of relatively simple functional forms to *ab initio* calculations, was rigorously shown to exhibit an LLT [28], confirming previous suggestive simulations [24]. Together, this body of work suggests the existence of an LLT in multiple families of empirical water models of progressively increasing complexity. However, strictly nonempirical (i.e., purely predictive) computational evidence of water's LLT has heretofore been lacking due to the significant increase in the computational cost of *ab initio* methods relative to empirical force fields [29].

Recently, a revolution in molecular modeling has begun to bridge this gap, namely the use of machine learning (ML) models trained to efficiently represent the potential energy surface (PES) predicted by computationally demanding first principles reference calculations [30-32]. These approaches significantly accelerate *ab initio* molecular dynamics simulations by performing the atomic energy and force calculations via a much-less-expensive surrogate ML model, rather than performing a full electronic structure calculation at each timestep. These ML models represent many-body correlations and capture polarizability effects present in the PES derived from electronic structure methods, and they have been successfully used to push the boundaries of problems accessible from first principles for a vast array of materials and fluids, including the properties and phase behavior of water [26,33-37]. In this work, we use one such ML simulation method, Deep Potential Molecular Dynamics (DPMD) [38,39], with a neural network model for water [33] trained on density functional theory (DFT) calculations with the Strongly-Constrained and Appropriately-Normed (SCAN) [40] exchange correlation functional. SCAN has been shown to be a leading semilocal functional in terms of providing a qualitatively accurate description of water's properties [41]. We previously demonstrated that a SCAN-based DPMD model exhibits physical properties consistent with the presence of an LLCP in water and provided a rough estimate for the LLCP location [26]. An updated and expanded version of this model was recently shown to successfully capture the equilibrium phase diagram of water's condensed phases, including high-pressure superionic ice states [33]. Herein, we sample water's metastable supercooled liquid state as predicted by the latest DPMD-SCAN model [33] using standard molecular dynamics simulations and two complementary enhanced sampling methods. Our results definitively show that this *ab initio* model exhibits an LLT and an LLCP at supercooled temperatures and positive pressures.

Fig. 1 shows trajectories of mass density ($\rho$) versus time for isothermal-isobaric MD simulations of metastable supercooled liquid water at temperature $T = 235$ K and pressure $P =$ 3000, 3200, and 3400 bar. At low $P$, two trajectories initialized from different high- and low-$\rho$ starting configurations converge to a low-$\rho$ state. At high $P$, the trajectories converge to a high-$\rho$ final state. But, at intermediate $P$, we observe long-lived high- and low-$\rho$ states, with reversible transitions between them. Supplemental Material (SM) Fig. S1 [42] confirms liquid-like structural relaxation at these conditions with no evidence of crystallization, though we do note a large heterogeneity in structural relaxation time in the low-$\rho$ state. Regardless, the lifetimes of the high- and low-$\rho$ liquids in a given trajectory are of a similar order (for LDL) or significantly longer (for HDL) than the associated relaxation time(s) for that trajectory. SM Fig. S2 [42] shows structural characterization of the high- and low-$\rho$ states at $T = 235$ K, $P = 3200$ bar, exhibiting oxygen-oxygen radial distribution functions consistent with the HDL and LDL phases reported with empirical molecular water models [23], and bimodal probability distributions in $\rho$ and the pairwise contribution to the excess entropy, $s_2$ (see Methods in the SM [42] for details on the $s_2$ calculation). Such behavior is consistent with a first-order phase transition along the $T = 235$ K isotherm, with LDL-HDL phase coexistence near $P = 3200$ bar.

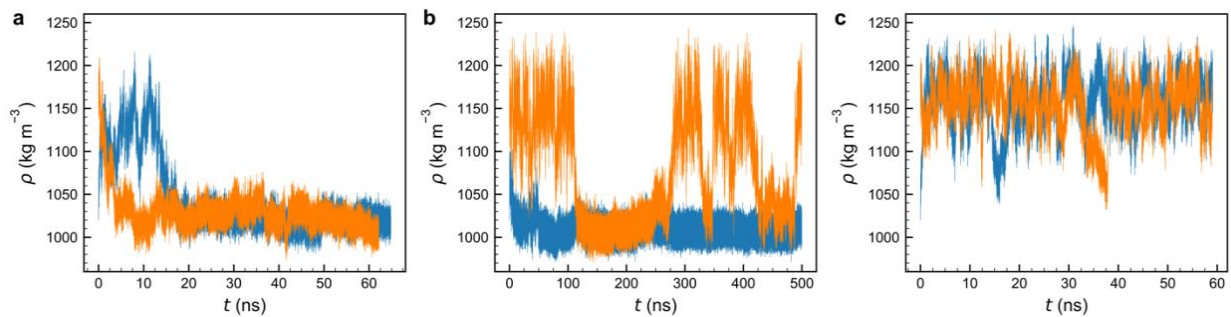

**Fig. 1:** Density from isothermal-isobaric simulations. Mass density ($\rho$) vs. time ($t$) trajectories for simulations with $N = 192$ molecules at $T = 235$ K and (a) $P = 3000$ bar, (b) $P = 3200$ bar, and (c) $P = 3400$ bar. In each panel, the orange trajectory was initialized in a high-$\rho$ configuration, and the blue trajectory was initialized in a low-$\rho$ configuration.

To confirm these results and provide rigorous thermodynamic evidence for an LLT, we performed umbrella sampling (US) simulations along the $s_2$ order parameter. $s_2$ is a function of the local structure of the fluid, and as shown in SM Fig. S2 [42], correlates closely with the density. Our hypothesis was that the local nature of $s_2$ could help drive the structural transformation between LDL and HDL more efficiently than a global order parameter (such as $\rho$). In Fig. 2, we plot free energy surfaces in the ($\rho$, $s_2$) space obtained from US simulations at the same conditions reported in Fig. 1. Confirming the results from the isothermal-isobaric simulations, following the $T$ = 235 K isotherm we observe a single dominant basin at (low-$\rho$, low-$s_2$) for low $P$, a dominant basin at (high-$\rho$, high-$s_2$) for high $P$, and two basins in approximate phase coexistence (equal free energy) at intermediate $P$. The right subfigure in each panel shows the free energy surface averaged over all $s_2$ values, demonstrating that the uncertainty in the free energy is ~1 $k_B T$ for all conditions, where $k_B$ is the Boltzmann constant. Crucially, SM Fig. S3 [42] shows that the locations of the (low-$\rho$, low-$s_2$) and (high-$\rho$, high-$s_2$) basins obtained via US agree very closely with the set of ($\rho$, $s_2$) values visited by the unbiased simulations reported in Fig. 1, lending credence to the validity of both results. Figs. 1 and 2 indicate a discontinuous first-order phase transition from LDL to HDL as pressure increases along the $T$ = 235 K isotherm, with approximate coexistence near $P$ = 3200 bar. We note that, at this state point, the free energy barrier separating the LDL and HDL phases calculated from the projection of the free energy surface along $\rho$ and $s_2$ is relatively mild (~1.5 $k_B T$) and only moderately larger than the uncertainty. Thus, sampling more deeply subcritical conditions (i.e., lower temperatures) would allow us to explore more obviously separated LDL and HDL phases.

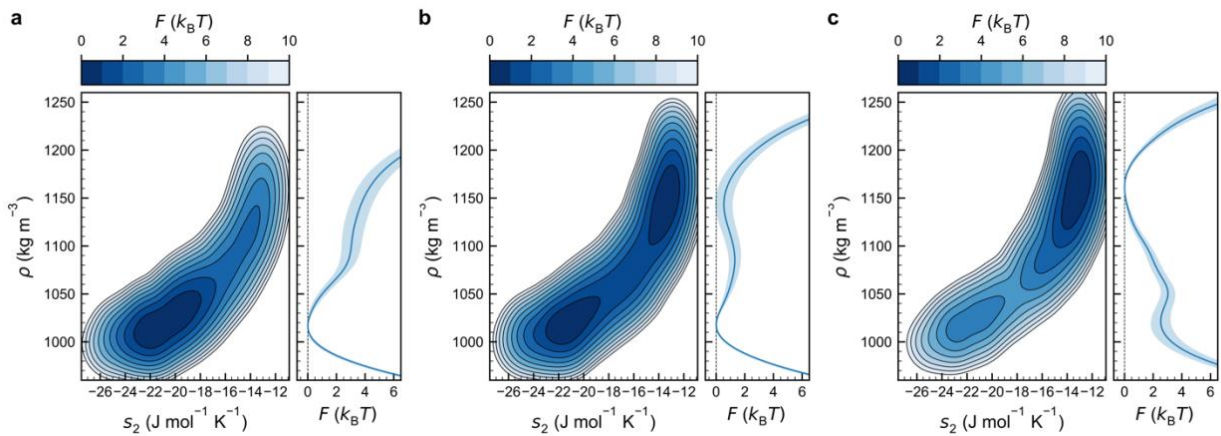

**Fig. 2:** Free energy surfaces from umbrella sampling (US) simulations. Free energy ($F$) as a function of $\rho$ and $s_2$ from US simulations with $N = 192$ molecules at $T = 235$ K and (a) $P = 3000$ bar, (b) $P = 3200$ bar, and (c) $P = 3400$ bar. In the left subfigure of each panel, contours represent 1 $k_BT$. In the right subfigure of each panel, the solid blue line represents $F$ averaged over all $s_2$ values, with the shaded regions representing 95% confidence intervals.

However, due to the slow structural relaxation times and significant computational expense of these simulations (~100x slower than classical empirical models), performing standard simulations at such low temperatures is computationally prohibitive at present. Thus, we turned to another advanced simulation technique, multithermal-multibaric (MTMB) sampling [43], in which biased sampling and histogram reweighting techniques enable exploration of a wide range of temperatures and pressures from only a single simulation. In the MTMB approach, results can be reweighted to any ($T, P$), provided that configurations relevant to that state point are appropriately sampled during the simulation. Thus, in our case, simulations can be performed at a nominally higher temperature (where thermalization of the liquid is easier to achieve) and reweighted to provide results at low-$T$. SM Fig. S4 [42] shows free energy surfaces obtained by MTMB run at a nominal temperature of $T = 280$ K and then reweighted to the same set of state points explored in Figs. 1 and 2 (see Methods), which demonstrate near-quantitative agreement between all three methods. Fig. 3a shows the free energy surface reweighted down to $T = 225$ K and $P = 3525$ bar, which has an LDL basin near $\rho = 1015$ kg m$^{-3}$ and $s_2 = -24$ J mol$^{-1}$ K$^{-1}$, and an HDL basin near $\rho$

= 1170 kg m$^{-3}$ and $s_2$ = -14 J mol$^{-1}$ K$^{-1}$. In Fig. 3b we show free energy surfaces from MTMB as a function of $\rho$ for a set of (T, P) that exhibit HDL and LDL basins at equal free energy (i.e., phase coexistence). As expected, the free energy barrier for the transition grows with decreasing temperature, reaching ~4 $k_B T$ at T = 225 K. Furthermore, this set of (T, P) defines the binodal line for the LLT, which we plot as a solid line in Fig. 3c.

One may approximately locate the LLCP as the (T, P) along the binodal at which the free energy barrier between HDL and LDL disappears. Based on the shape and location of the binodal, we locate the critical temperature and pressure as $T_c$ = 242 ± 5 K and $P_c$ = 2950 ± 150 bar, indicated by the shaded region in Fig. 3c. We note that due to the uncertainty in the free energy surfaces (~1 $k_B T$), this approach only provides an estimate of the critical temperature and pressure. However, free energy surfaces calculated along isotherms and isobars sufficiently far from the critical point (SM Fig. S5 [42]) show a continuous crossover from low to high $\rho$ at supercritical temperatures or low pressures and a discontinuous transition at subcritical temperatures or high pressures. This result demonstrates definitively that an LLCP does exist in this model, somewhere in the region 225 K < T < 245 K and 2750 bar < P < 3250 bar. The extension of the binodal line to supercritical conditions is the Widom line, which we illustrate via the locus of maximum isothermal compressibility (dashed line in Fig. 3c), which extends from the critical point to higher T and lower P.

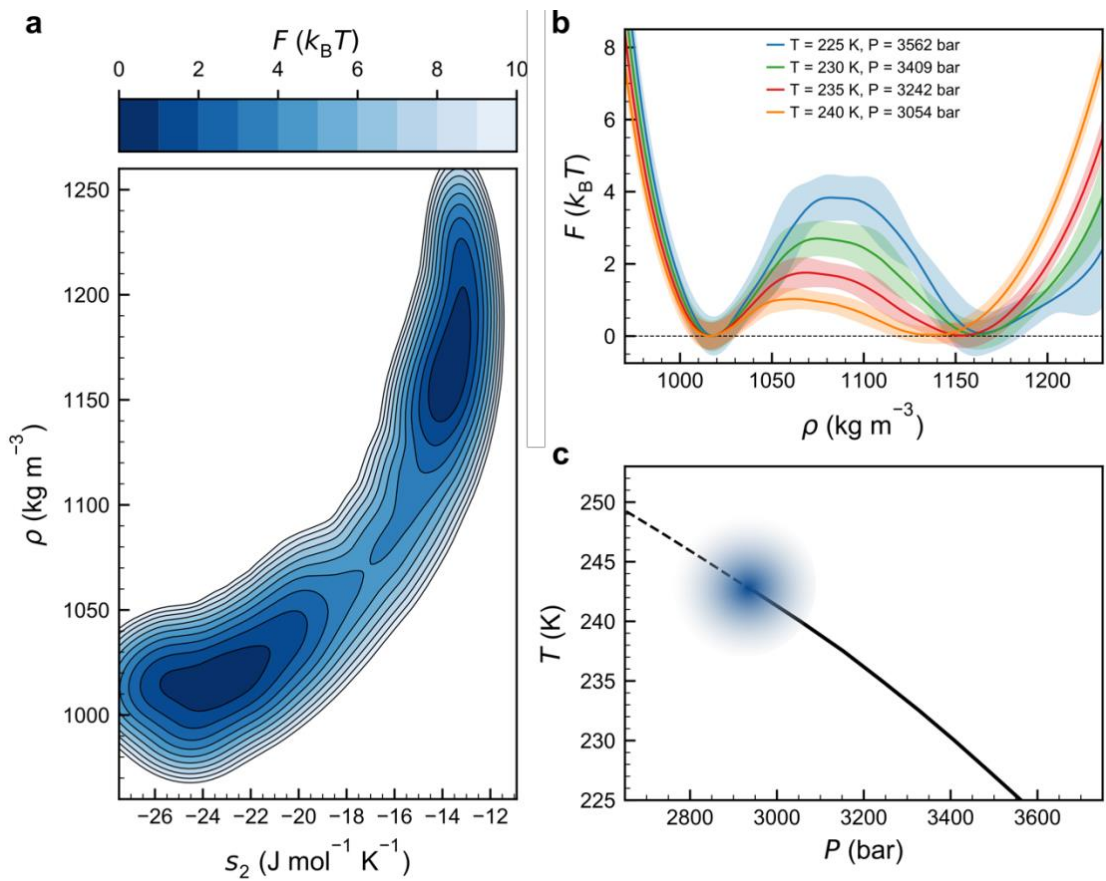

**Fig. 3:** Phase coexistence from multithermal-multibaric (MTMB) simulations. (a) Free energy ($F$) as a function of $\rho$ and $s_2$ from MTMB simulations with $N = 192$ molecules at $T = 225$ K and $P = 3525$ bar. (b) $F$ vs. $\rho$ at various $T$ and $P$ as marked. Shaded regions represent 95% confidence intervals. (c) Liquid-liquid binodal (solid black line), approximate critical point location (shaded blue region), and line of maximum isothermal compressibility (dashed black line) in the $T$-$P$ plane.

These results provide strong computational evidence that liquid water exhibits a metastable LLT and LLCP at supercooled temperatures (i.e., well below the melting temperature of ice predicted from SCAN [33,34]) and positive pressures. Three separate sampling methods show results in near-quantitative agreement, and two different free energy estimates rigorously show liquid-liquid coexistence with a discontinuous transition in density under sufficiently supercooled conditions. We note that our estimate for the critical point location ($T_c = 242 \pm 5$ K and $P_c = 2950 \pm 150$ bar) is at somewhat higher $T$ and $P$ than recent estimates from available experimental data [44,45]. However, SCAN is known to overestimate the strength of water's hydrogen bond [33,46], which may stabilize the LDL phase with respect to HDL due to LDL's highly tetrahedral local structure, and correspondingly shift the LDL-HDL phase boundary to higher $T$ and $P$. With further developments in computing power and simulation algorithms it may be quite instructive to perform a similar study with a ML model trained on higher levels of theory, such as DFT based on a hybrid functional, coupled cluster methods, etc. Similarly, evaluating potential system size effects (including verifying the expected scaling of the LDL-HDL free energy barrier with system size [22]) would also be a worthwhile avenue for future work.

We emphasize that this study was performed with a slightly different model than our prior work on the LLCP in a DPMD-SCAN model [26], the training set herein being expanded to include additional configurations at high temperatures and pressures [33]. The critical point location reported in our prior study ($T_c$ = 224 ± 3 K, $P_c$ = 2687 ± 68 bar) [26] is at somewhat lower $T$ and $P$ than our present estimate. Some portion of this discrepancy may come from differences in the DPMD training dataset between the two model versions. However, our prior work also used an equation-of-state approach that was fit to simulation data collected at supercritical conditions and then extrapolated to provide information at low $T$. That approach, while helpful in providing an initial estimate for the critical point location in a computationally-efficient manner, may not be quantitatively accurate due to the extrapolation needed to reach near-critical conditions. We argue that the present work, which instead samples the LLT directly, should be considered the more definitive estimate for the LLCP location predicted by SCAN. Nevertheless, it is important to note that in both studies, the training dataset did not explicitly include any configurations specifically targeted toward the LLT, only the various ordered solid phases and liquid water across a wide range of ($T$, $P$). We posit that the fact that both versions of the model exhibit physical properties consistent with an LLT/LLCP, despite only being trained on water's equilibrium ice and liquid phases, makes the present results qualitatively robust to the particularities of the model's training dataset, provided that the ML model sufficiently reproduces SCAN's potential energy surface. In other words, the presence of the LLCP has significant effects even at ($T$, $P$) far away from criticality, such that properties learned at state points away from the LLCP still encode its existence in the ML model.

Our results represent some of the strongest computational evidence to date for water's LLT, as they were obtained from nonempirical (purely *ab initio*) approaches. Apart from adding to our current understanding of water's physical chemistry, we anticipate that similar approaches (ML-based *ab initio* models combined with enhanced sampling techniques) will provide fertile ground to push the boundaries of computational physics/chemistry for other fluids and materials, as has been recently demonstrated for the LLT in phosphorous [47].

**Methods, Data, and Code Availability:**

Detailed methods are provided in the SM [42], which includes Refs. [48-58]. All data and code related to this work, including simulation input files, raw simulation trajectory data, analysis scripts, and processed data used to create all figures in the manuscript, are available for download at the Princeton DataSpace repository [59].

**Acknowledgements:**

The work of T.E.G., P.M.P, R.C., P.G.D. and A.Z.P. was supported by the "Chemistry in Solution and at Interfaces" (CSI) Center funded by the U.S. Department of Energy Award DE-SC001934. Computational resources were managed by Princeton Research Computing, a consortium of groups including the Princeton Institute for Computational Science and Engineering (PICSciE) and the Office of Information Technology's High Performance Computing Center and Visualization Laboratory at Princeton University.

**Author contributions:**

P.G.D., A.Z.P., and T.E.G. conceived of the project; T.E.G and P.M.P. performed research; all authors designed research, discussed results, and wrote the paper.

**Supplemental Material for:**

**Liquid-liquid transition in water from first principles**

**Authors:**

Thomas E. Gartner III[1,†], Pablo M. Piaggi[1], Roberto Car[1,2,3,4], Athanassios Z. Panagiotopoulos[5,*], and Pablo G. Debenedetti[5,*]

[1]Department of Chemistry, Princeton University, Princeton, NJ 08544;

[2]Department of Physics, Princeton University, Princeton, NJ 08544;

[3]Program in Applied and Computational Mathematics, Princeton University, Princeton, NJ 08544;

[4]Princeton Institute for the Science and Technology of Materials, Princeton University, Princeton, NJ 08544;

[5]Department of Chemical and Biological Engineering, Princeton University, Princeton, NJ 08544

[†]Current address: School of Chemical and Biomolecular Engineering, Georgia Institute of Technology, Atlanta, GA 30332

[*]Corresponding authors: azp@princeton.edu, pdebene@princeton.edu

**Methods:**

We performed classical molecular dynamics simulations using the Deep Potential Molecular Dynamics [1,2] method (DeepMD-kit v1.0) interfaced with the LAMMPS [3] simulation software (v3Mar20), using the SCAN [4]-based ML model for water developed in Ref. [5]. For all simulations, we used a system size of 192 water molecules in a cubic periodic simulation box. We maintained temperature and pressure using the LAMMPS Nosé-Hoover-type thermostat and barostat, with relaxation times 50 fs and 500 fs, respectively. Since static thermodynamic properties in classical simulations do not depend on the mass of the atoms, we used a hydrogen mass of 2 AMU in the time integration to enable a time step size of 0.5 fs, but used water's average molar mass of 18.015 g mol$^{-1}$ for analysis (e.g., computation of the mass density). The standard isothermal-isobaric simulations were run for 60-500 ns depending on the state point.

We confirmed the accuracy of the ML model in representing the SCAN PES for deeply supercooled water by selecting 302 configurations representative of the HDL and LDL states from the isothermal-isobaric simulations at $T = 235$ K and $P = 3200$ bar and recomputing the atomic energies and forces from density functional theory (DFT) calculations (see SM Fig. S6). We performed the DFT calculations with the SCAN functional using the Quantum Espresso package version 7.0 [6,7] augmented by the LIBXC 5.2.2 library [8]. We employed norm-conserving, scalar-relativistic pseudopotentials for O and H parameterized using the PBE functional [9]. Kinetic energy cutoffs of 110 Ry and 440 Ry were used for the wavefunctions and the charge density, respectively. Only the Γ point of the Brillouin zone was sampled and the convergence absolute error for the self-consistent procedure was set to $10^{-6}$ Ry. All other parameters were set to their default values in Quantum Espresso. The average energy error between the ML model and SCAN DFT was 1.2 meV per H$_2$O, and the average force error was 99 meV Å$^{-1}$. The magnitude of the errors is comparable with that of other state-of-the-art ML potentials. We note a slight systematic shift in the relative energy of HDL vs. LDL configurations, in which the ML model slightly overpredicts the potential energy difference between HDL and LDL relative to SCAN (see SM Fig. S6c). This systematic error is likely due to the absence of deeply supercooled liquid configurations in the ML model training data set. While such errors may influence the precise location of phase coexistence in the ($T$, $P$) plane due to slight differences in the relative stability of the liquid phases, we do not anticipate the overall qualitative conclusions of the study to be affected.

We used LAMMPS patched with the development version of the PLUMED software [10] to perform the US and MTMB biased sampling simulations. For the US simulations, we performed 12 separate simulations with the centers of harmonic bias potentials located at -26.6, -24.9, -23.3, -21.6, -19.9, -19.1, -18.3, -17.5, -16.6, -15.0, -13.3, and -11.6 J mol$^{-1}$ K$^{-1}$ (SM Fig. S7a) with a spring constant of 200 $k_B^{-2}$ kJ mol$^{-1}$, with $k_B$ the Boltzmann constant. The order parameter $s_2$ is the pairwise contribution to the excess entropy [11], defined as a function of the oxygen-oxygen radial distribution function $g(r)$.

$$s_2 = -2\pi\rho k_B \int_0^{r_{max}}[g(r) \ln g(r) - g(r) + 1]r^2 dr \qquad (1)$$

Although this formula is strictly valid for the upper limit of integration ($r_{max}$) tending to infinity, for practical reasons one must define a finite $r_{max}$. In this work, we chose $r_{max} = 5$ Å, which resulted in well-separated values of $s_2$ for the LDL and HDL states while limiting the computational expense of the $s_2$ calculation (SM Fig. S2b). For each US simulation, we performed 10 ns of equilibration followed by 20 ns of sampling with two independent replicates for the $P = 3000$ and $P = 3400$ state points and three replicates for the $P = 3200$ state point. SM Fig. S7b shows that these simulation lengths are longer than the structural relaxation timescale for all except the lowest $s_2$ values. However, the free energy surface obtained excluding the most sluggish simulations (SM Fig. S7c) is very similar to that reported in main text Fig. 2b. All initial configurations for the US simulations were selected randomly from the MTMB trajectory. Free energy surfaces from US were calculated using the binless Weighted Histogram Analysis Method (WHAM) [12], with uncertainties calculated as the 95% confidence interval obtained from a block-bootstrap approach [13] using blocks of length 1 ns and 200 bootstrap samples.

We performed the MTMB simulation [14] using the on-the-fly probability enhanced sampling (OPES) framework [15,16], biasing the $\rho$ and potential energy ($U$) using the ECV_MULTITHERMAL_MULTIBARIC action in the PLUMED OPES module and biasing $s_2$ using the ECV_UMBRELLAS_LINE action. The bias potential along $s_2$ was composed of the sum of gaussian umbrellas of width SIGMA = 0.8 J mol$^{-1}$ K$^{-1}$ spaced equally from -29.1 < $s_2$ < -10.8 J mol$^{-1}$ K$^{-1}$ with a BARRIER setting of 20 $k_BT$. Four independent walkers were used to speed the exploration of the free energy surface and improve sampling (see the trajectories in SM Fig. S8). The MTMB simulation was run for 400 ns at a nominal $T$ = 280 K and $P$ = 2000 bar, sampling the range 220 K < $T$ < 285 K and 500 bar < $P$ < 4000 bar. The bias potential was updated every 200 time steps, and all other parameters were set to their default values in PLUMED. Fluid properties and free energy surfaces from the MTMB simulation (ignoring the first 10 ns of the trajectory while the bias potential converged) were calculated using reweighting techniques as described in Ref. [15], with uncertainties calculated as the 95% confidence interval obtained from a block-bootstrap approach using blocks of length 2.4 ns and 200 bootstrap samples.

All LAMMPS and PLUMED input scripts, as well as Python scripts used for analysis and plotting, are posted on our associated data repository [17].

**Supplemental Figures:**

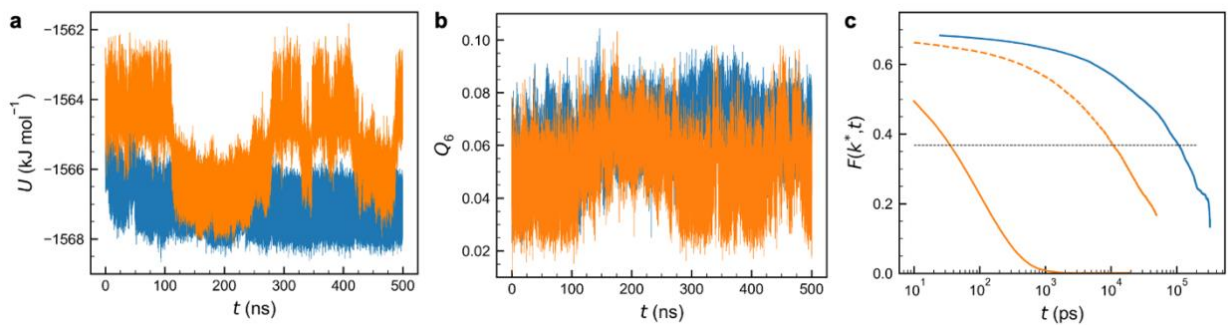

**SM Fig. S1:** Additional data from isothermal-isobaric simulations at $T$ = 235 K and $P$ = 3200 bar. (a) Potential energy ($U$) and (b) global rotationally-invariant bond orientational order parameter ($Q_6$) vs. time ($t$) trajectories. The $Q_6$ was calculated as described in the PLUMED documentation (https://www.plumed.org/doc-master/user-doc/html/_q6.html). The orange trajectory was initialized in a high-$\rho$ configuration, and the blue trajectory was initialized in a low-$\rho$ configuration. (c) Self-part of the intermediate scattering function, $F(k^*; t)$, at the wavenumber of maximum intensity in the structure factor, $k^*$. The solid orange line was computed over 20 ns < $t$ < 100 ns from the orange trajectory (characteristic of HDL), the dashed orange line was computed over 130 ns < $t$ < 230 ns from the orange trajectory (characteristic of LDL) and the blue line was computed over 50 ns < $t$ < 370 ns from the blue trajectory (characteristic of LDL). The black dashed line denotes a value of 1/$e$, a convenient metric for the structural relaxation time in the system.

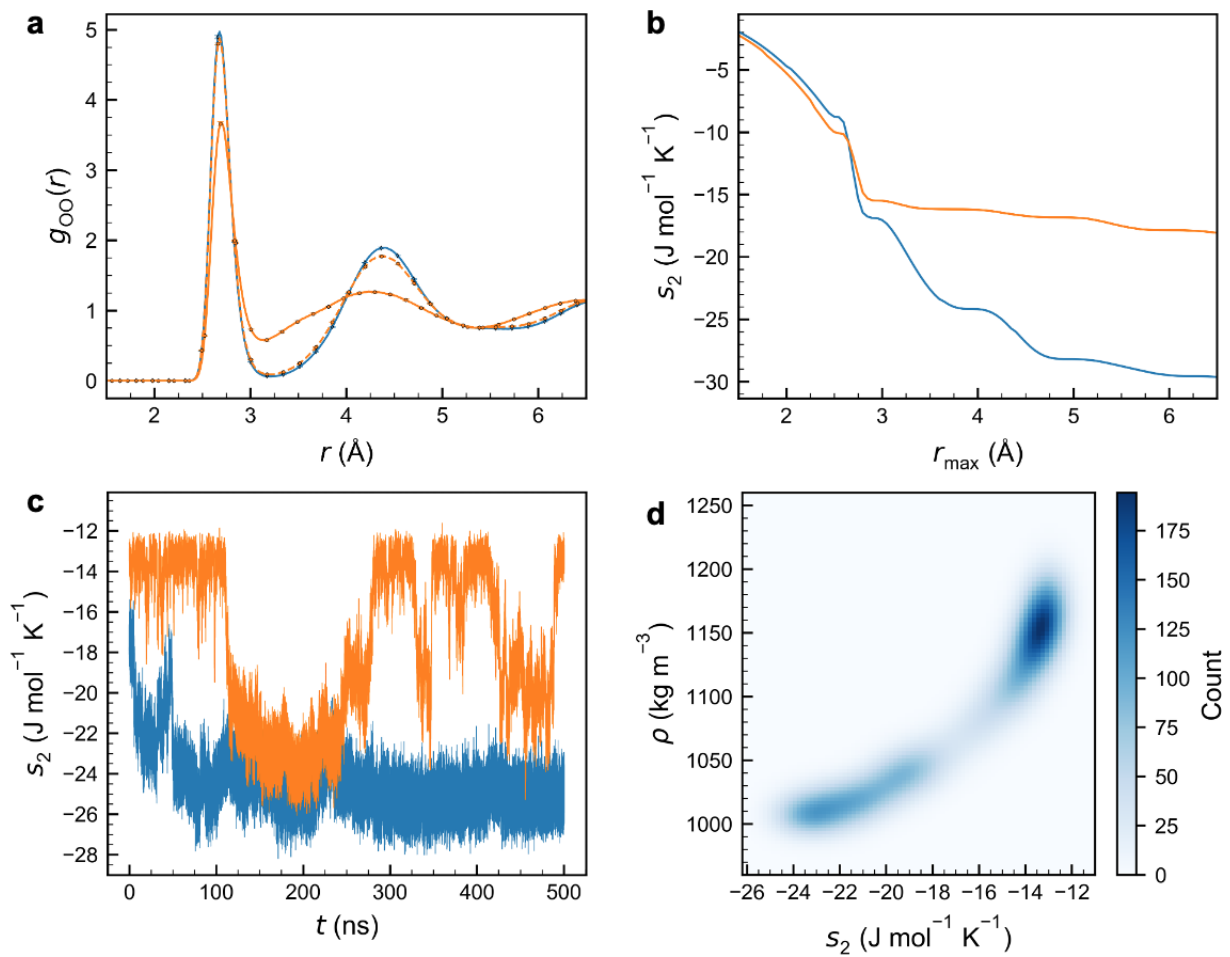

**SM Fig. S2:** HDL and LDL structural characterization at $T = 235$ K and $P = 3200$ bar. (a) Oxygen-oxygen radial distribution function ($g_{OO}(r)$) from isothermal-isobaric simulations, computed from the same portions of the trajectories as described in SM Fig. S1c. (b) Variation in the pairwise contribution to the excess entropy ($s_2$) as a function of the maximum radius considered in the calculation ($r_{max}$) for HDL configurations (orange) and LDL configurations (blue). (c) Trajectory of $s_2$ vs. $t$ for the simulations reported in main text Fig. 1b. (d) 2-dimensional histogram of $\rho$ and $s_2$ for the orange trajectory.

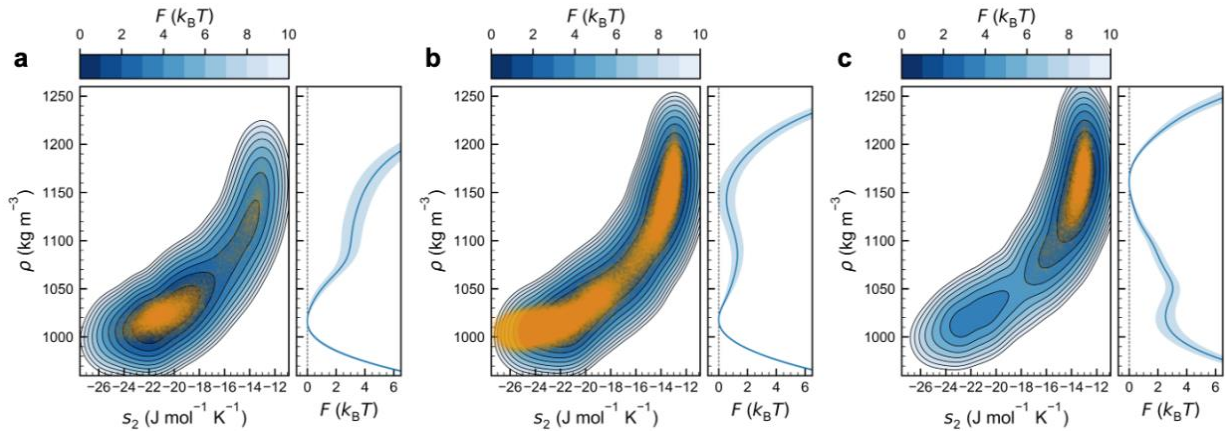

**SM Fig. S3:** Comparison of isothermal-isobaric and US simulations. (a-c) Same as the free energy surfaces reported in main text Fig. 2, with the configurations visited in the isothermal-isobaric simulations reported in main text Fig. 1 superimposed as yellow points.

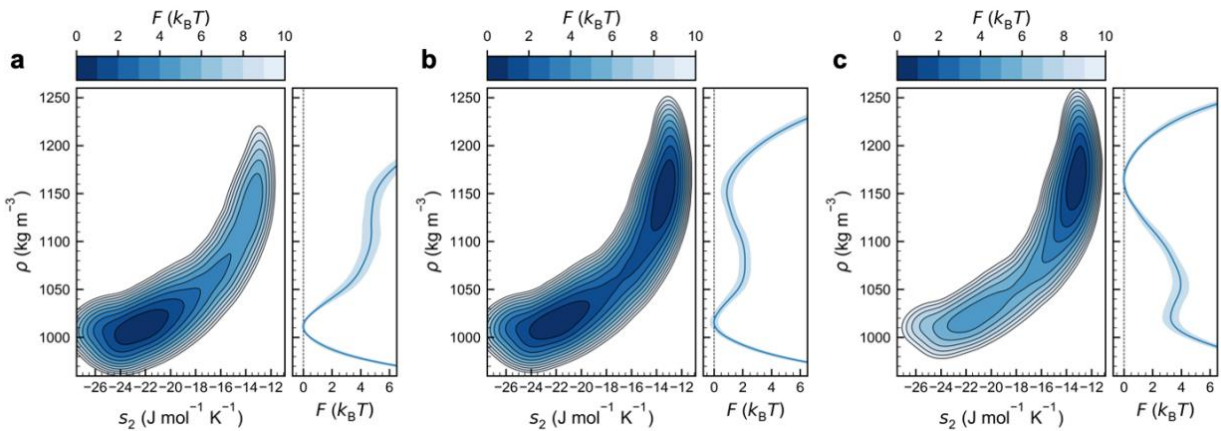

**SM Fig. S4:** Free energy surfaces from MTMB simulation. (a-c) Free energy surfaces obtained from reweighting the MTMB simulation to the same set of state points reported in main text Fig. 2.

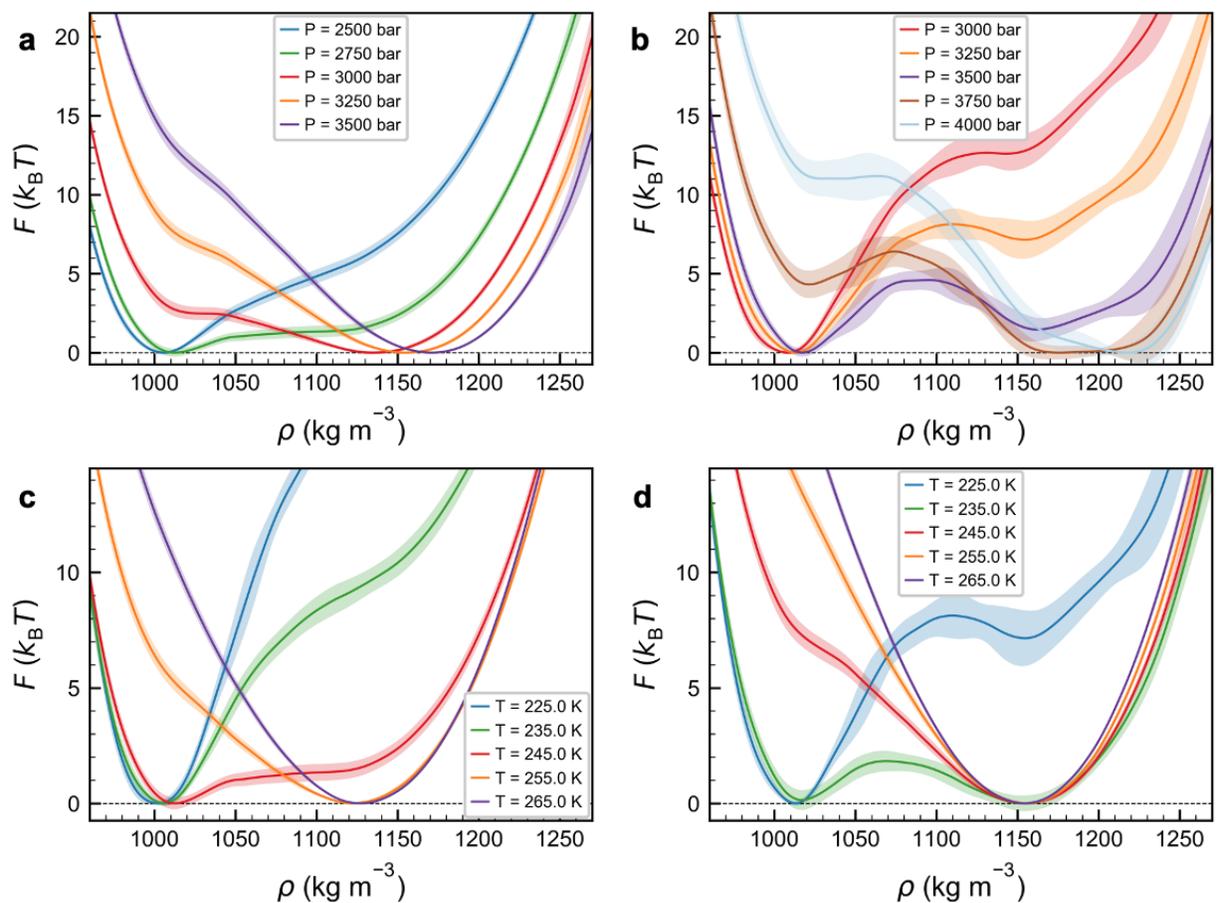

**SM Fig. S5:** Transitions along isotherms and isobars. Free energy ($F$) as a function of $\rho$, obtained from reweighting the MTMB simulation to various state points along (a) the $T = 245$ K isotherm, (b) the $T = 225$ K isotherm, (c) the $P = 2750$ bar isobar, and (d) the $P = 3250$ bar isobar.

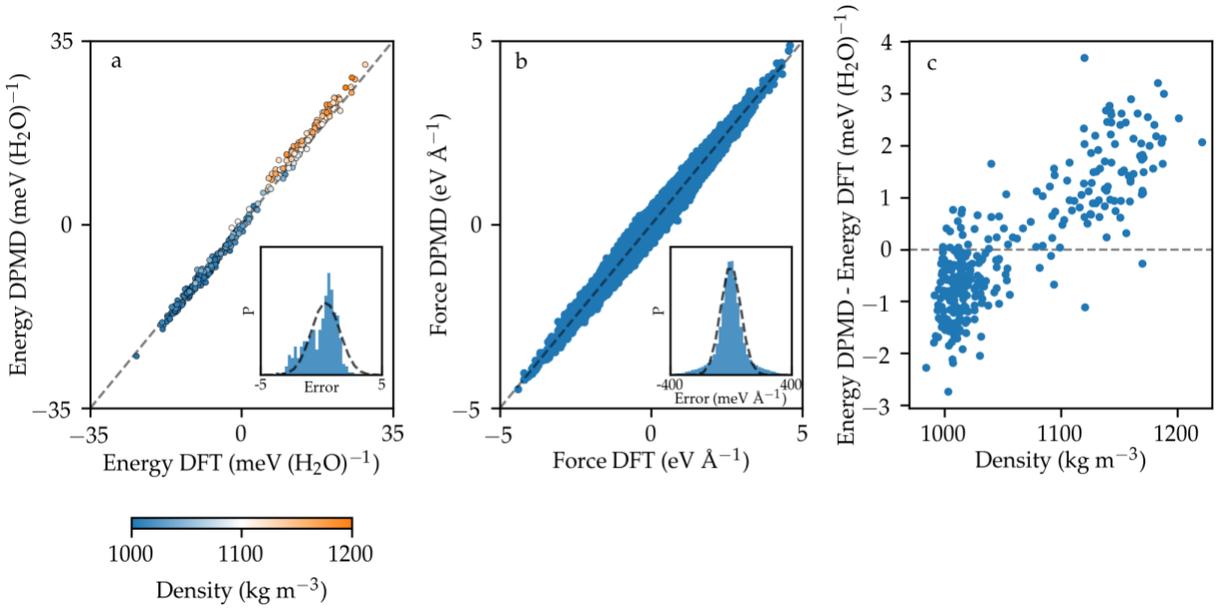

**SM Fig. S6:** Comparison between ML potential (DPMD) and DFT for configurations representative of the LDL and HDL states. (a) Potential energy calculated using DFT calculations and the ML potential. (b) Forces evaluated using DFT and the ML potential. The insets in (a) and (b) show the distribution of errors defined as the difference in energies and forces, respectively, between the ML potential and DFT. (c) Difference in potential energy calculated using the ML potential and DFT vs. mass density.

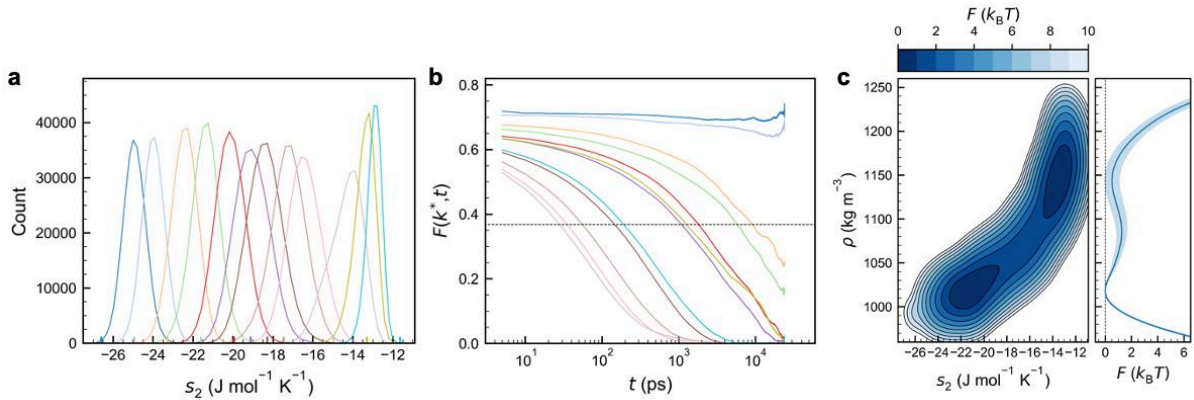

**SM Fig. S7:** Additional information about US simulations. (a) Histogram of $s_2$ values visited by the US simulations at $T = 235$ K and $P = 3200$ bar. Each color represents a different umbrella window, with the center of the window marked by a small tick mark along the x-axis. (b) Decay of $F(k^*; t)$ for the same set of simulations reported in (a). The black dashed line denotes a value of $1/e$, a convenient metric for the structural relaxation time in the system. (c) Free energy surface obtained from US in the same fashion as main text Fig. 2b, but with the data from the two most sluggish US simulations (dark blue and light blue lines in (a) and (b)) removed from the calculation.

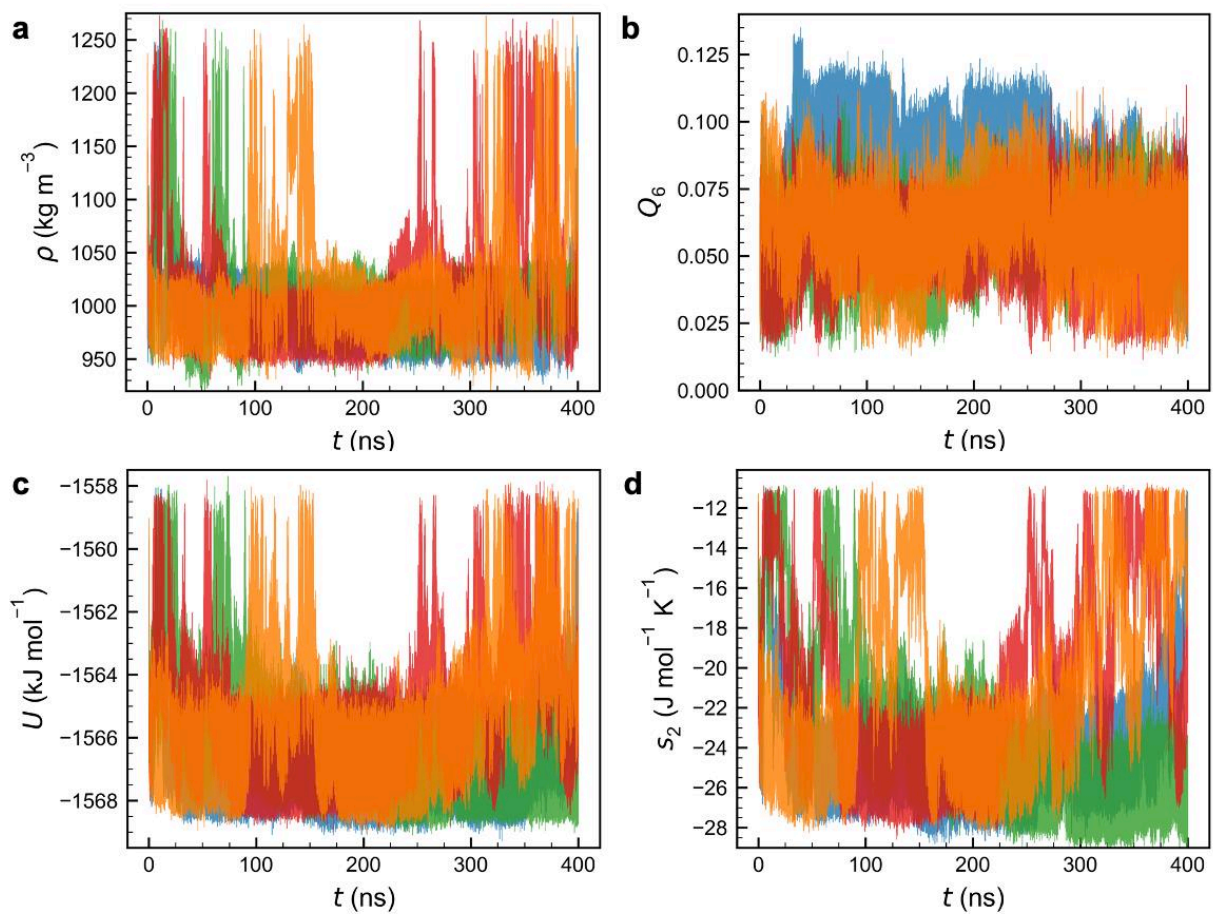

**SM Fig. S8:** Time trajectories from MTMB simulation. (a) $\rho$, (b) $Q_6$, (c) $U$, and (d) $s_2$ versus $t$ for the MTMB simulation. Colors indicate 4 independent walkers.

**Supplemental References:**